\documentclass[epj]{webofc}
\usepackage[utf8]{inputenc}
\usepackage[varg]{txfonts}   
\usepackage{booktabs}
\usepackage{xcolor}
\definecolor{darkred}{rgb}{0.4,0.0,0.0}
\definecolor{darkgreen}{rgb}{0.0,0.4,0.0}
\definecolor{darkblue}{rgb}{0.0,0.0,0.4}
\usepackage[bookmarks,linktocpage,colorlinks,
    linkcolor = darkred,
    urlcolor  = darkblue,
    citecolor = darkgreen]{hyperref}
\wocname{EPJ Web of Conferences}
\woctitle{Lattice2017}
\newcommand{\NN}{\ensuremath{\mathcal{N}}}
\newcommand{\NNc}{\ensuremath{\mathcal{N}^c}}
\newcommand{\NNt}{\ensuremath{\widetilde{\mathcal{N}^c}}}
\begin{document}
%
\selectlanguage{english}
\title{%
ChPT loops for the lattice: pion mass and decay constant, HVP at finite volume and $n\bar n$-oscillations\fnsep\thanks{Presented at Lattice 2017, the 35th International Symposium
  on Lattice Field Theory, Granada, Spain, 18-24 June 2017}
}
\author{%
\firstname{Johan} \lastname{Bijnens}\inst{1}\fnsep\thanks{\email{bijnens@thep.lu.se}}
}
\institute{%
Department of Astronomy and Theoretical Physics, Lund University, S\"olvegatan 14A, SE22362 Lund, Sweden}
\abstract{%
I present higher loop order results for several calculations in
Chiral perturbation Theory.
1) Two-loop results at finite volume for hadronic vacuum
polarization. 2) A three-loop calculation of the pion mass and
decay constant in two-flavour ChPT. For the pion mass all needed auxiliary
parameters can be determined from lattice calculations of $\pi\pi$-scattering.
3) Chiral corrections to neutron-anti-neutron oscillations.}
\begin{flushright}
LU TP 17-32\\
October 2017
\end{flushright}
\vskip-1cm
\maketitle
\section{Introduction}\label{intro}

This talk presents a number of calculations done using Chiral Perturbation
Theory (ChPT) that should be useful for lattice calculations.
The three parts that will be discussed are the vector-two-point function
at two-loop order including partial quenching, twisting and finite volume.
A preliminary version of the paper \cite{BijnensYYY} can be found in the
thesis by Johan Relefors~\cite{thesisRelefors}. 
The second part is about the first full
three-loop calculation in mesonic ChPT, the pion mass and decay constant
in the two-flavour case~\cite{Bijnens:2017wba}. The third part is about the
construction of ChPT operators for neutron-antineutron oscillations and
the one-loop calculation of chiral and finite volume
corrections~\cite{BijnensXXX,thesisKofoed}. These three parts are independent
of each other, the common ground is that they all use ChPT.
An introduction to ChPT for lattice practitioners is \cite{Golterman:2009kw}.

\section{The vector two-point function and HVP}
\label{sec:HVP}

This work was done in collaboration with Johan Relefors. The main reason
to consider this quantity
is that the lowest order hadronic vacuum polarization contribution to the muon
anomaly $a_\mu=(g_\mu-2)/2$ can be obtained from
\begin{align}
 a_\mu^{\mathrm{LO,HVP}}=\,&
 \int_0^\infty dQ^2 f\left(Q^2\right)\left[\Pi^{(1)}_{ee}\left(Q^2\right)-\Pi^{(1)}_{ee}\left(0\right)\right]\,.
\end{align}
The function $f(Q^2)$ is well known. The two-point function of vector currents
is:
\begin{align}
\Pi^{\mu\nu}_{ab}(q) \equiv\,& i\int d^4x e^{iq\cdot x}
\big<T(j^\mu_a(x)j_a^{\nu\dagger}(0))\big>\,
&\Pi^{\mu\nu}_{ab} =\,& \left(q^\mu q^\nu - q^2 g^{\mu\nu}\right)
\Pi^{(1)}_{ab}\,.
\end{align}
The last equation is valid in infinite volume for the conserved currents we use here:
\begin{align}
j_{\pi^+}^\mu =\,& \bar d\gamma^\mu u\,~
&j_{u}^\mu =\,& \bar u\gamma^\mu u,~
&j_{d}^\mu =\,& \bar d\gamma^\mu d,~
&j_{s}^\mu =\,& \bar s\gamma^\mu s,~
&j_{e}^\mu =\,& (2/3)\bar u\gamma^\mu u-(1/3)\bar d\gamma^\mu d
\left(-(1/3)\bar s\gamma^\mu s\right)\,.
\end{align}

\subsection{Connected versus disconnected}
\label{sec:disconnected}

\begin{figure}
\sidecaption
\includegraphics[width=0.5\textwidth]{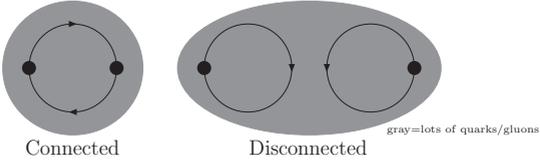}
\caption{Connected (left) and disconnected (right) diagram(s) for the
two-point function. Lines are valence quarks in a sea of quarks and gluons.
Figure from \cite{Bijnens:2016ndo}.}
\label{figdisconnected}
\end{figure}

In lattice calculations there is a connected and disconnected part shown
schematically in figure~\ref{figdisconnected}.
The disconnected part is often more difficult to calculate on the lattice so
an analytic understanding of the relative sizes is very useful. This was
done in ChPT at one-loop in \cite{DellaMorte:2010aq} where a ratio of $1/10$
was found in two-flavour ChPT. In \cite{Bijnens:2016ndo} the argument was
extended to two-loop order. At that order contributions from singlet vector
current operators start contributing that break the ratio of $1/10$, however
loops with the singlet current only start at even higher order.
The latter is the reason for the factor $1/10$ as explained
in~\cite{Bijnens:2016ndo} and in \cite{Bijnens:2016pci}.
An estimate of that contribution using vector-meson-dominance (VMD) and
the two-loop calculations gave reasonable agreement with the lattice results,
many of which can be found in these proceedings.

\begin{figure}
\begin{minipage}{0.33\textwidth}
\includegraphics[width=0.99\textwidth]{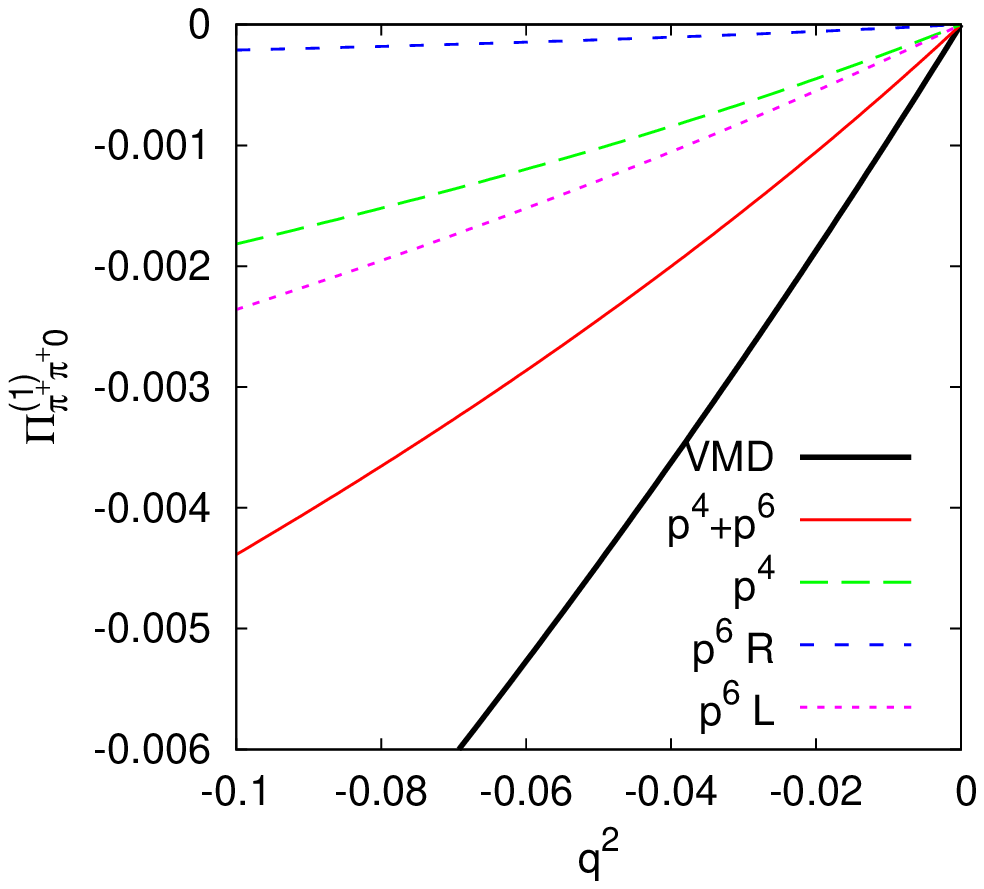}\\[-3mm]
\centerline{(a)}
\end{minipage}
\begin{minipage}{0.33\textwidth}
\includegraphics[width=0.99\textwidth]{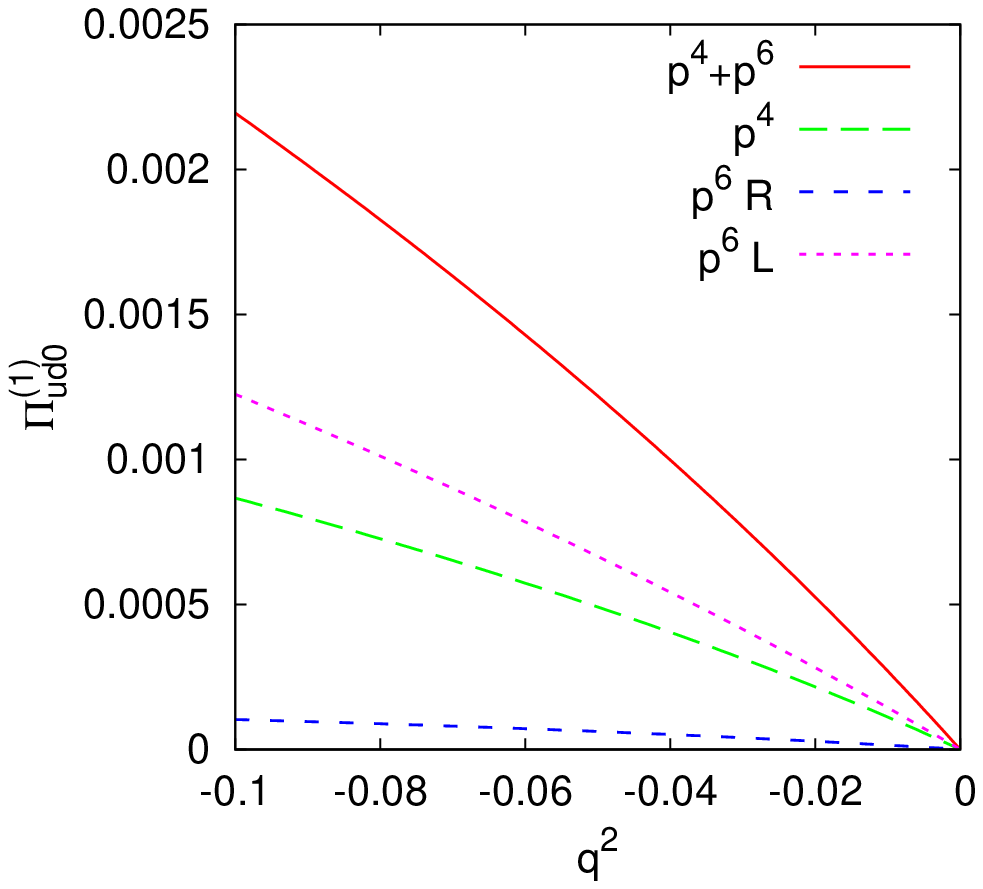}\\[-3mm]
\centerline{(b)}
\end{minipage}
\begin{minipage}{0.33\textwidth}
\includegraphics[width=0.99\textwidth]{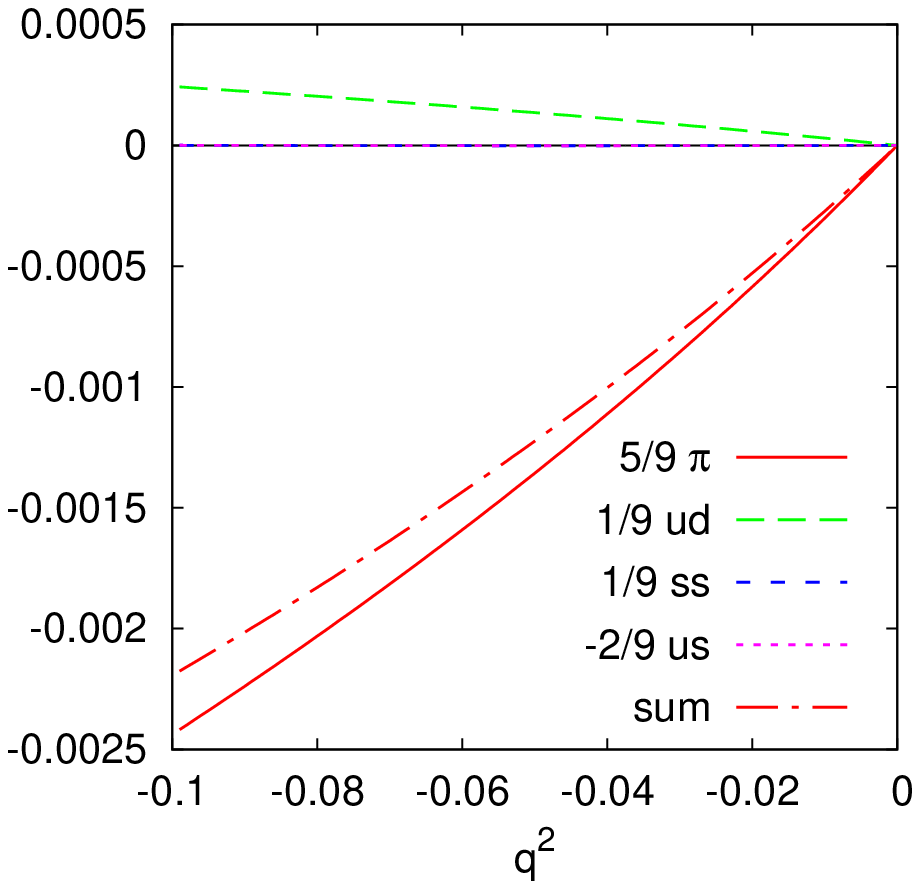}\\[-3mm]
\centerline{(c)}
\end{minipage}
\caption{The vector two-point functions. (a) The disconnected part $\hat\Pi^{(1)}_{\pi^+\pi^+0}$ (b) The disconnected part $\hat\Pi^{(1)}_{ud0}$ (c)~Various contributions to the electromagnetic $\hat\Pi^{(1)}_{ee}$.
Figures from \cite{Bijnens:2016ndo}.}
\label{fignumdisconnected}
\end{figure}

We define the subtracted quantity
\begin{align}
\Pi^{(1)}_{ab0}(q^2) =\,& \Pi^{(1)}_{ab}(q^2)-\Pi^{(1)}_{ab}(0)\,.
\end{align}
$ab=\pi^+\pi^+$ gives the connected contribution for a single quark current
and $ab=ud$ the disconnected part. The electromagnetic (two-flavour) case
is given by $\hat\Pi^{(1)}_{ee}=(5/9)\hat\Pi^{(1)}_{\pi^+\pi^+}+(1/9)\hat\Pi^{(1)}_{ud}$. The results are shown in figure~\ref{fignumdisconnected}.
The connected part shown in (a) has the VMD part as the largest contribution,
but loops at $p^6$ are larger than those at $p^4$. This is especially due to the
diagrams involving $L_9^r$. The disconnected part shown in (b) is for the loops
with pions exactly $-1/2$ the connected part, as followed from the
two-flavour singlet argument.
There are small corrections from Kaon loops. In (c) we show the parts
for the electromagnetic case. One can see that the strange quark current
contributions are really small. A comparison with lattice data can be found in
\cite{Bijnens:2016ndo}. 

\subsection{Twisting and finite volume}

\begin{figure}
\sidecaption
\begin{minipage}[t]{0.33\textwidth}
\includegraphics[width=0.99\textwidth]{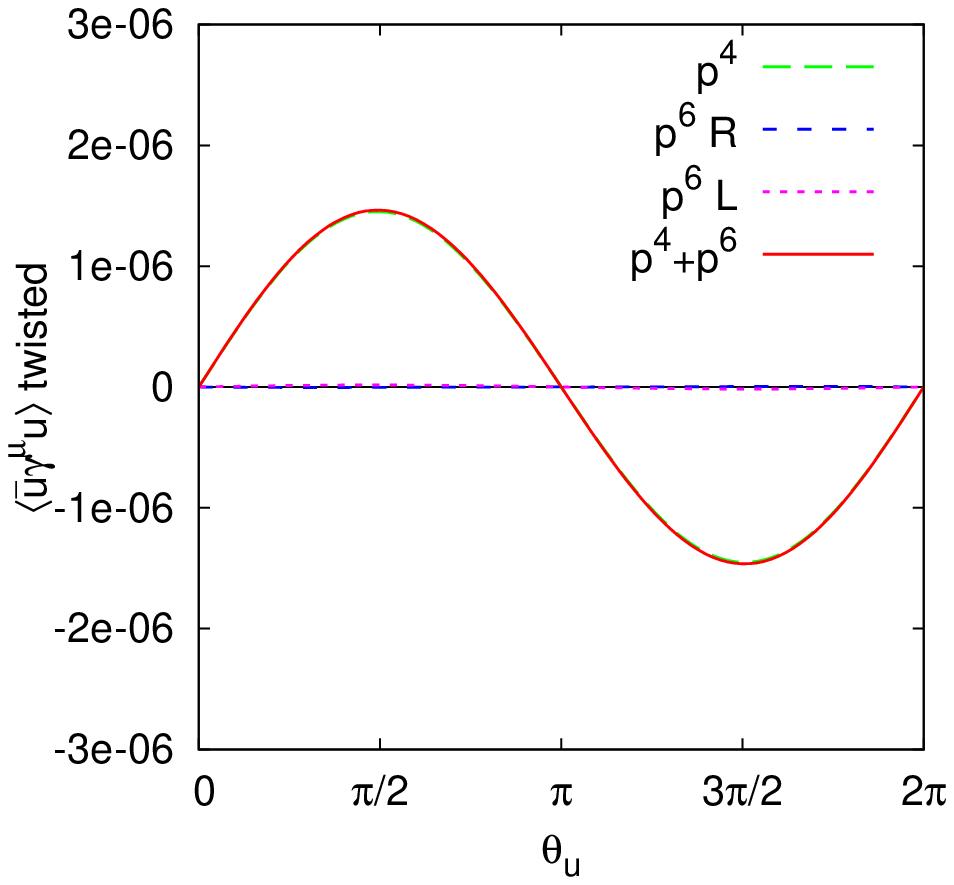}\\[-3mm]
\centerline{(a)}
\end{minipage}
\begin{minipage}[t]{0.33\textwidth}
\includegraphics[width=0.99\textwidth]{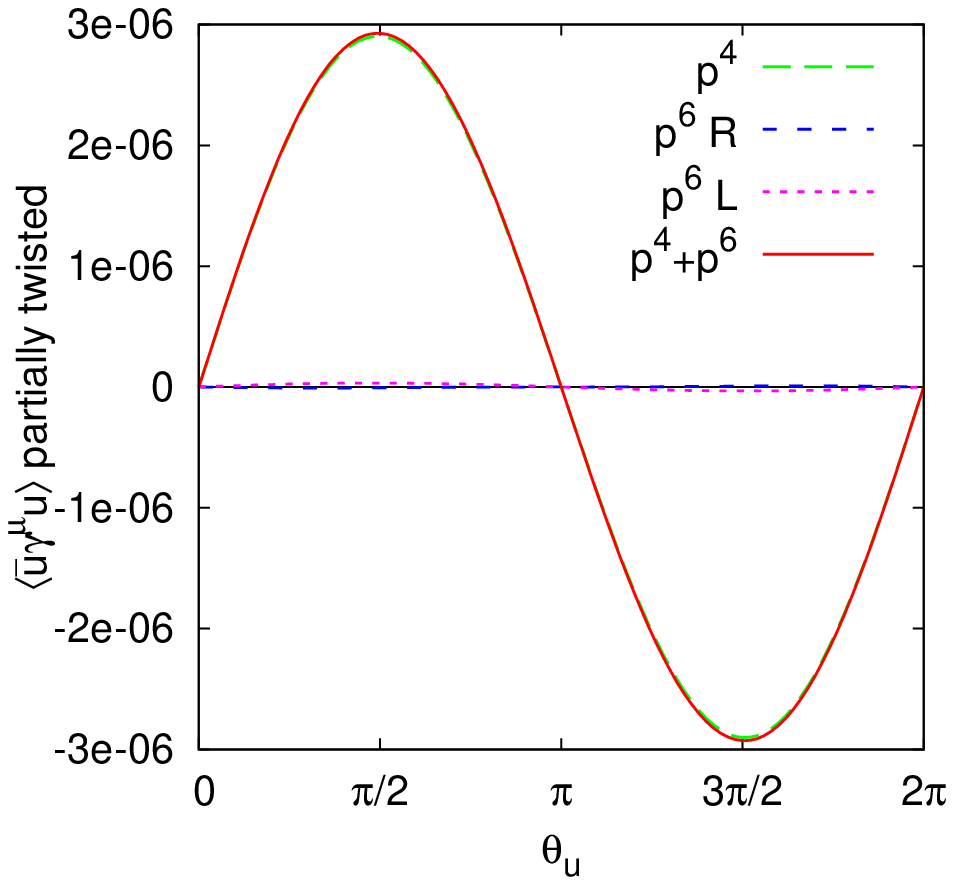}\\[-3mm]
\centerline{(b)}
\end{minipage}
\caption{The vector vacuum expectation value as a function of twist angle.
(a) Fully (b) Partially twisted.}
\label{figtwistvev}
\end{figure}

The finite volume calculation at one-loop was done
in~\cite{Aubin:2013daa,Aubin:2015rzx} and found to agree
well with lattice data. Here we discuss the extension to two-loop order.
Twisted boundary conditions are useful since on a lattice with periodic
boundary conditions momenta are $p^i=2\pi n^i/L$ with $n^i$ integer.
Only a few different values of low $q^2$ are thus directly accessible.
Putting a constraint on a quark field in some directions
$q(x^i+L)= \exp(i\theta^i_q)q(x^i)$ changes the momenta to
$p^i=\theta^i/L+2\pi n^i/L$ allowing many more $q^2$. This can also be
done only for the valence quarks (partial twisting). However, this twisting
changes the Ward identities \cite{Aubin:2013daa,Bijnens:2014yya}.
The underlying current is still conserved but now the vector current can
have a vacuum expectation value changing the Ward identity to
\begin{align}
q_\mu\Pi^{\mu\nu}_{\pi^+\pi^+}=\,&\left\langle\bar u\gamma^\nu u-\bar d\gamma^\nu d
\right\rangle\,.
\end{align}
Partially quenched and twisted ChPT at one and two-loop satisfy this.
The numerical size of the effect is quite small~\cite{BijnensYYY}.
The vacuum expectation
value as a function of the twist angle for a fully and partially twisted
up-quark with a twist angle in one direction only is shown in figure~\ref{figtwistvev}. These are for $m_\pi L=4$, the $p^6$ corrections are really small.
The numbers should be compared with the scalar vacuum expectation value
and its finite volume correction~\cite{BijnensYYY,Bijnens:2006ve},
$\langle \bar u u\rangle~~\approx -1.2~10^{-2}~\mathrm{GeV}^{3}$ and
$\langle \bar u u\rangle^V\approx -2.4~10^{-5}~\mathrm{GeV}^{3}$.

\begin{figure}
\sidecaption
\begin{minipage}[t]{0.33\textwidth}
\includegraphics[width=0.99\textwidth]{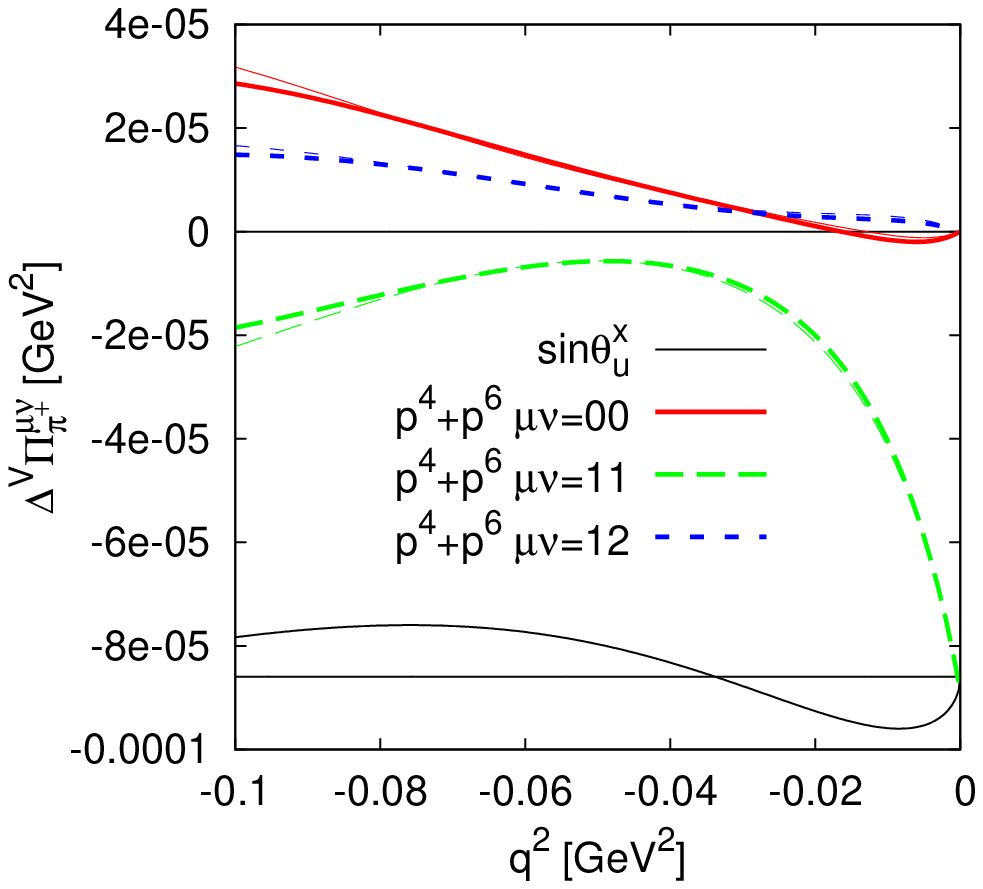}\\[-3mm]
\centerline{(a)}
\end{minipage}
\begin{minipage}[t]{0.33\textwidth}
\includegraphics[width=0.99\textwidth]{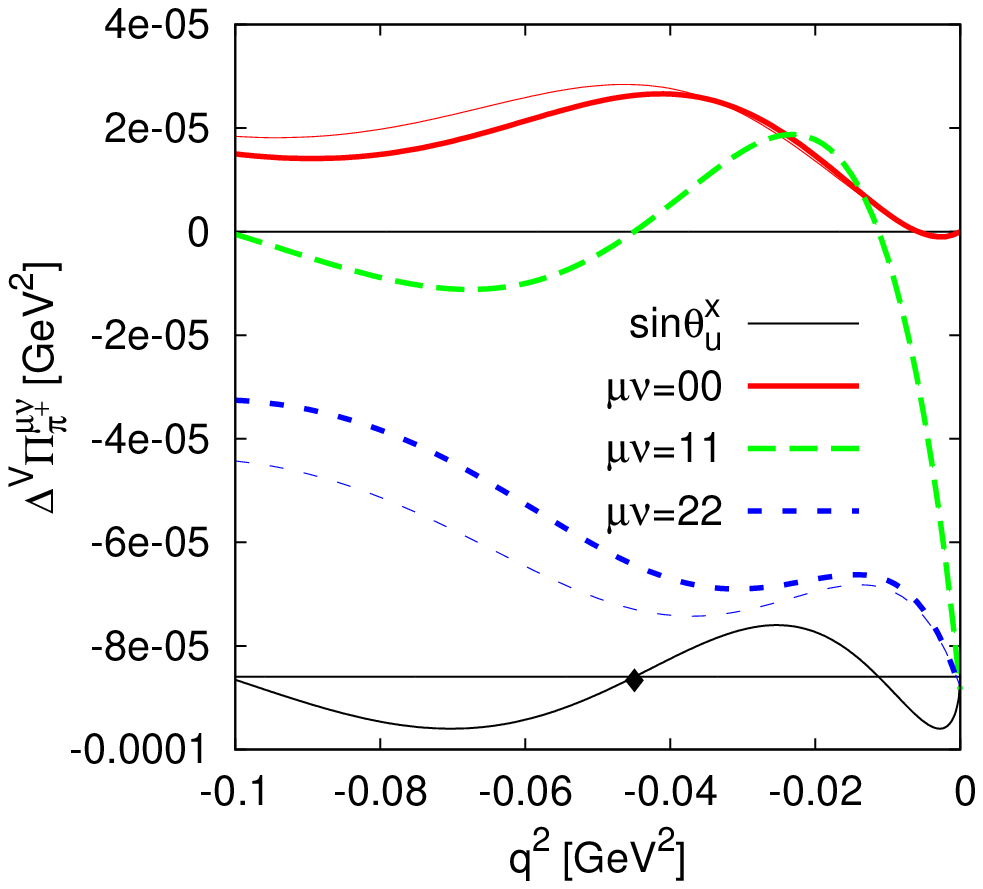}\\[-3mm]
\centerline{(b)}
\end{minipage}
\caption{Finite volume corrections to some components of the two-point
function $\Pi^{\mu\nu}_{\pi^+\pi^+}$. The thin lines are $p^4$-only, the thick
lines $p^4+p^6$. (a) Spatially symmetric twisting (b) twisting in one 
direction only. The diamond indicates a $q^2$ accessible with periodic
boundary conditions. Bottom curve shows $\sin\theta^x_u$, the sine of the twisting angle in the $x$-direction, they help understanding the shape of the curves. Figures from \cite{BijnensYYY}.
}
\label{figtwistpi}
\end{figure}

How large are now the corrections due to finite volume and twisting?
The low-energy constants (LECs) we use are those of \cite{Bijnens:2014lea}.
The calculations were done in \cite{BijnensYYY}. Numerical results
for the finite volume corrections at $m_\pi L=4$ as a function of $q^2$
are shown in figure~\ref{figtwistpi}. These should be compared to
the VMD contribution which is of order $0.005~(q^2/0.1~\mathrm{GeV}^2)$.
So the finite volume correction are rather $q^2$ dependent but not large.
In particular, the $p^6$-corrections are very small, very different from
the infinite volume result. One can also see the difference between different
ways of obtaining the same $q^2$ by partial twisting, in (a) it is done in
a spatially
symmetric fashion, in (b) only in one direction. The difference allows
for checking the finite volume corrections with the same underlying lattice.

\begin{figure}
\sidecaption
\begin{minipage}[t]{0.33\textwidth}
\includegraphics[width=0.99\textwidth]{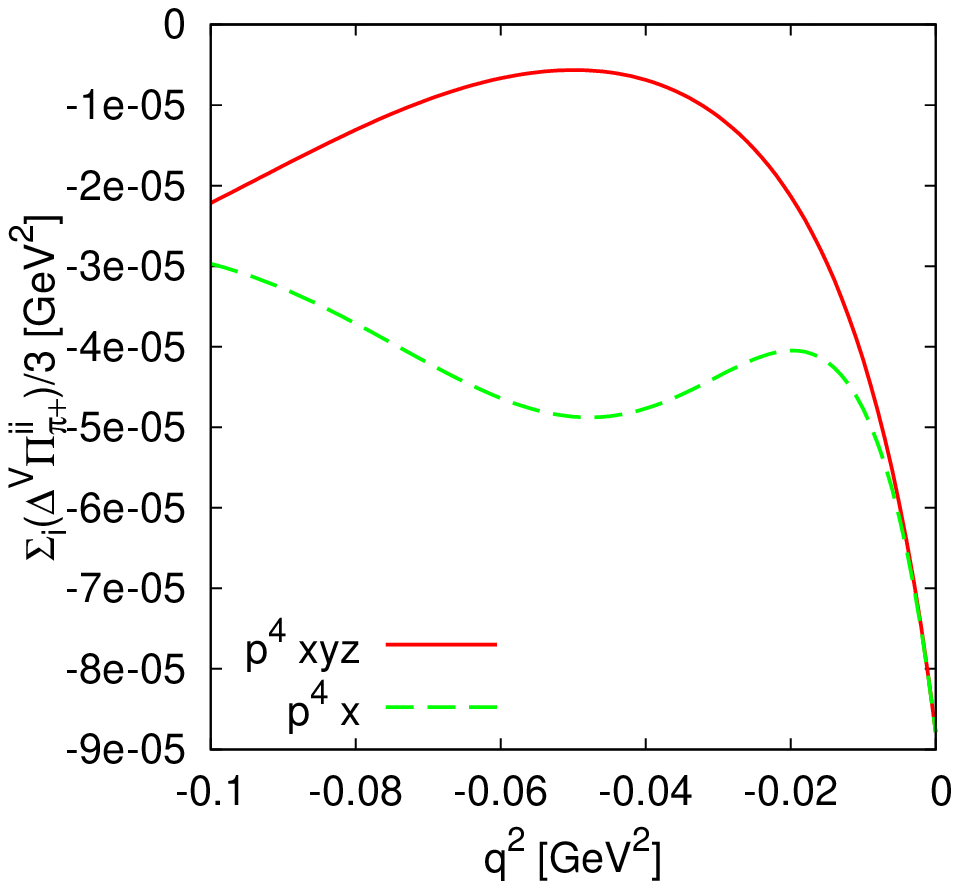}\\[-3mm]
\centerline{(a)}
\end{minipage}
\begin{minipage}[t]{0.33\textwidth}
\includegraphics[width=0.99\textwidth]{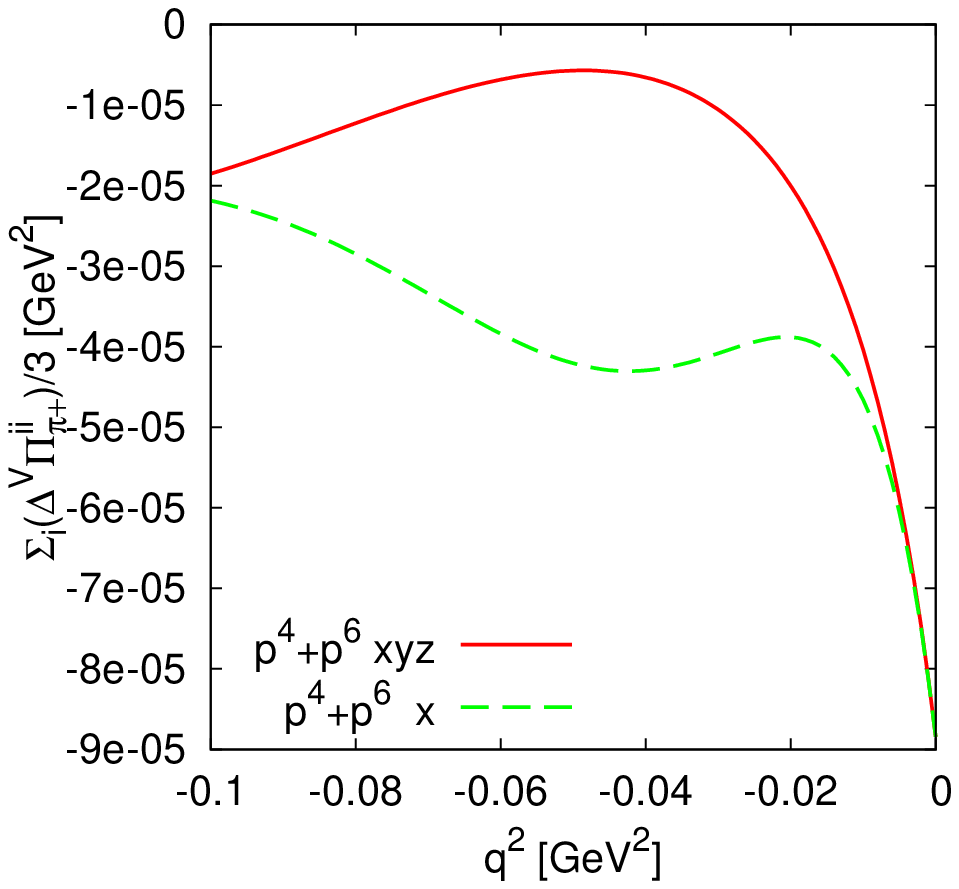}\\[-3mm]
\centerline{(b)}
\end{minipage}
\caption{Finite volume corrections to the trace over spatial components $\overline\Pi= (1/3)\sum_{i=x,y,z}\Pi^{ii}$ as a function of $q^2$ for two different ways of twisting, spatially symmetric and in one direction only. (a) $p^4$ (b) $p^4+p^6$. Figures from \cite{BijnensYYY}.}
\label{figtwistpibar}
\end{figure}

One often calculates on the lattice the spatial average over the two-point
function components. This is defined in the caption of figure~\ref{figtwistpibar} and numerical results at order $p^4$ are shown in figure~\ref{figtwistpibar}(a) and the full $p^4+p^6$ result in figure~\ref{figtwistpibar}(b). Again,
the difference between the two can be used to see check the estimates of
the finite volume corrections using only one underlying lattice.

\section{The pion mass and decay constant at three-loops}
\label{sec:pion}

This is work done in collaboration with Nils Hermansson Truedsson~\cite{Bijnens:2017wba}. The pion mass and decay
constant have been calculated in ChPT before at tree-level \cite{GellMann:1968rz}, one-loop chiral logarithms \cite{Langacker:1973hh}, full one-loop
and incidentally the proper start of ChPT \cite{Gasser:1983yg}, as well as
two-loop \cite{Burgi:1996qi,Bijnens:1995yn,Bijnens:1997vq}. Each new step
introduced a number of new methods. Lowest order (LO) and chiral logs were
done with current algebra. The full one-loop calculation was done by hand and by
directly expanding the functional integral with the help of \textsc{REDUCE}.
The NNLO or two-loop work was done by hand with a little help
from \textsc{FORM}. The NNNLO result required a large amount of use
of \textsc{FORM}~\cite{Vermaseren:2000nd}. The main stumbling block is really
the integrals. The reduction to master integrals was done with
\textsc{REDUZE}~\cite{Studerus}. All master integrals are
known~\cite{Laporta:1996mq,Melnikov:2000zc}.

\begin{figure}[thb] 
  \centering
  \sidecaption
  \includegraphics[width=0.5\textwidth,clip]{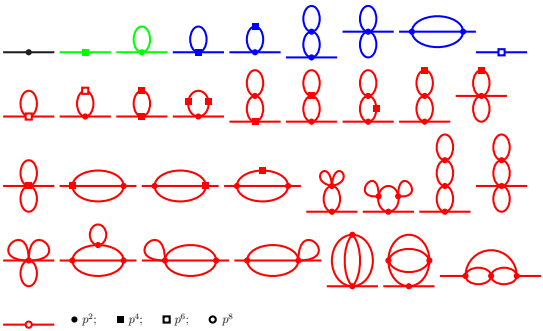}
  \caption{The diagrams for the pion mass and decay constant to NNNLO in ChPT. Figure from \cite{Bijnens:2017wba}.}
  \label{fig:massdiagrams}
\end{figure}

A calculation of this size also requires a large number of checks. The nonlocal
divergences must cancel as in any field theory. We used two different
parametrizations of the Lagrangians, square root and exponential, and the
leading logarithms agreed with the result derived earlier \cite{Bijnens:2009zi,Bijnens:2010xg}. The diagrams are shown in figure~\ref{fig:massdiagrams}.

We can do the expansion of the physical quantities $M_\pi^2,F_\pi$
in terms the LO quantities $M^2,F$ ($x$-expansion) or the inverse
($\xi$-expansion). $M^2 = 2B\hat m$ and $F$ are the LO pion mass and decay
constant. The $x$-expansions are
\begin{align}
\label{xParamEqn}
M_{\pi}^{2} =\,& M^{2}\Big\{ 1+x\left(a_{10}^{M}+a_{11}^{M}L_{M}\right) +x^{2}\left(a_{20}^{M}+a_{21}^{M}L_{M}+a_{22}^{M}L_{M}^{2}\right)
 +x^{3}\left(a_{30}^{M}+a_{31}^{M}L_{M}+a_{32}^{M}L_{M}^{2}+a_{33}^{M}L_{M}^{3}\right)
\Big\}
\nonumber\\
F_{\pi}=\,&F \Big\{ 1+x\left(a_{10}^{F}+a_{11}^{F}L_{M}\right) +x^{2}\left(a_{20}^{F}+a_{21}^{F}L_{M}+a_{22}^{F}L_{M}^{2}\right)
 +x^{3}\left(a_{30}^{F}+a_{31}^{F}L_{M}+a_{32}^{F}L_{M}^{2}+a_{33}^{F}L_{M}^{3}\right)
\Big\}
\end{align}
with $x=M^{2}/(16\pi ^{2}F^{2})$ and $L_{M} = \log( M^{2}/\mu ^{2})$, and
the $\xi$-expansions are
\begin{align}
\label{xiParamEqn}
M^{2} =\,& M_{\pi}^{2}\Big\{ 1+\xi \left(b_{10}^{M}+b_{11}^{M}L_{\pi}\right) +\xi ^{2}\left(b_{20}^{M}+b_{21}^{M}L_{\pi}+b_{22}^{M}L_{\pi}^{2}\right)
 +\xi ^{3}\left(b_{30}^{M}+b_{31}^{M}L_{\pi}+b_{32}^{M}L_{\pi}^{2}+b_{33}^{M}L_{\pi}^{3}\right)
\Big\}
\nonumber\\
F=\,&F_{\pi} \Big\{ 1+\xi \left(b_{10}^{F}+b_{11}^{F}L_{\pi }\right) +\xi ^{2}\left(b_{20}^{F}+b_{21}^{F}L_{\pi}+b_{22}^{F}L_{\pi}^{2}\right)
 +\xi ^{3}\left(b_{30}^{F}+b_{31}^{F}L_{\pi}+b_{32}^{F}L_{\pi}^{2}+b_{33}^{F}L_{\pi}^{3}\right)
\Big\}
\end{align}
where $\xi=M_{\pi}^{2}/(16\pi ^{2}F_{\pi}^{2})$ and $L_{\pi} = \log( M_{\pi}^{2}/\mu ^{2})$.
The analytical values for the coefficients can be
found in~\cite{Bijnens:2017wba}. We found that the coefficients of the
logarithms for the mass at order $p^8$ can all be written in terms of
coefficients obtainable from the
lattice from $\pi\pi$-scattering to two-loop order.

For the numerical estimates, we use $\mu = 0.77$~GeV,
 $\bar l_1 = -0.4, \bar l_2 = 4.3$, $\bar l_3=3.41,\bar l_4=4.51$
and the $\pi\pi$-scattering estimates from~\cite{Bijnens:1997vq}.
All remaining LECs have been set to zero. The resulting values for the
coefficients are give in table~\ref{NumParamsTable}.
\begin{table}
\sidecaption
 \begin{tabular}[b]{|c||c |c |c|c|} 
 \hline
 $ij$ & $a_{ij}^{M}$ & $b_{ij}^{M}$ & $a_{ij}^{F}$ & $b_{ij}^{F}$ \\ [0.7ex]
 \hline
 \hline
 10 & 0.0028 & $-$0.0028 & 1.0944 & $-$1.0944\\ 
 \hline
 11 & 0.5 & $-$0.5 & $-$1.0 &1.0\\
 \hline
 20 & 1.6530 &  $-$1.6577& $-$0.0473 & $-$1.1500\\
 \hline
 21 & 2.4573 & $-$3.2904 & $-$1.9058 & 4.1388\\
 \hline
 22 & 2.125 & $-$0.625 & $-$1.25 & $-$0.25\\ 
 \hline
  30 & 0.4133 & $-$6.8035 & $-$244.5350 & 242.2724 \\
 \hline
   31 &$-$3.7044 &  4.2718&  $-$15.4989& 28.5703\\
  \hline
    32 & 17.1476 & 0.6204& $-$9.3946 & $-$6.7751\\
\hline
 33 & 4.2917 &  5.1458&  $-$3.4583 & $-$0.4167\\
 \hline
\end{tabular}
\caption{Numerical values of the $a_{ij}^{M,F}$ and $b_{ij}^{M,F}$ for the input
parameters given in the text. Table from \cite{Bijnens:2017wba}.}
\label{NumParamsTable}
\end{table}
The numerical values of $a_{30}^F$ and $b_{30}^F$ are rather large,
due to a very large numerical coefficient $383293667/1555200\approx 246.5$
appearing. The remaining coefficients are of natural magnitude.

\begin{figure}[htb]
\begin{minipage}{0.33\textwidth}
\includegraphics[width=0.99\textwidth]{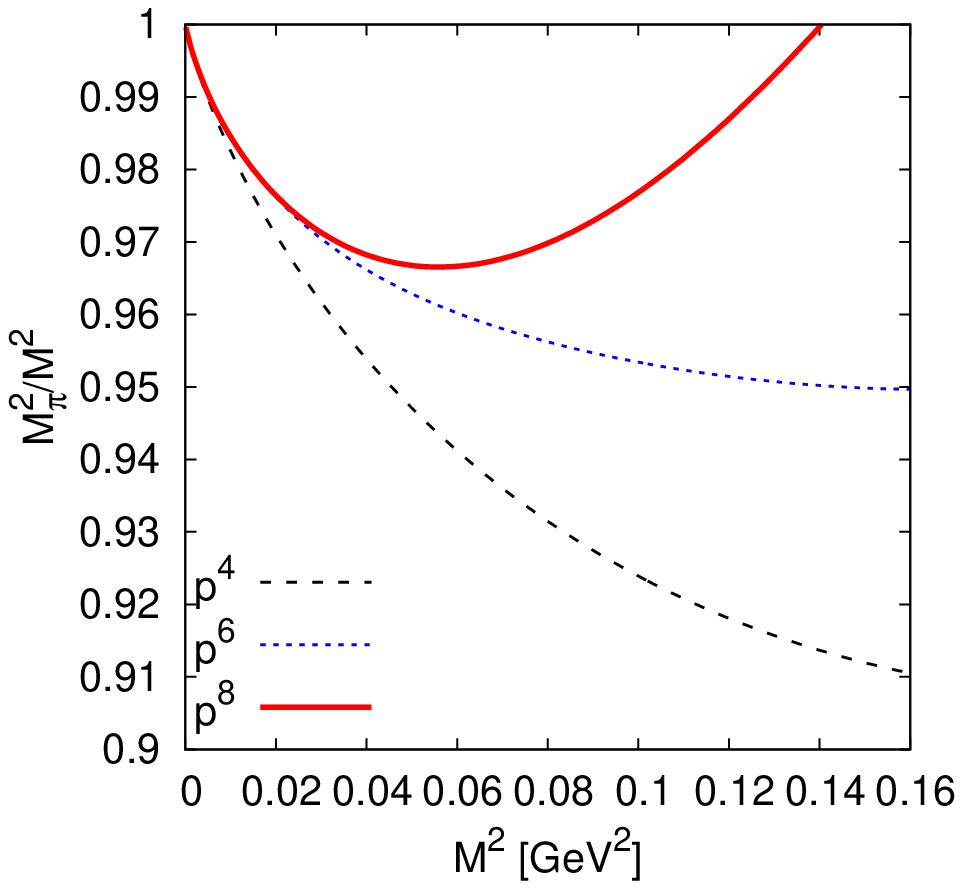}\\[-3mm]
\centerline{(a)}
\end{minipage}
\begin{minipage}{0.33\textwidth}
\includegraphics[width=0.99\textwidth]{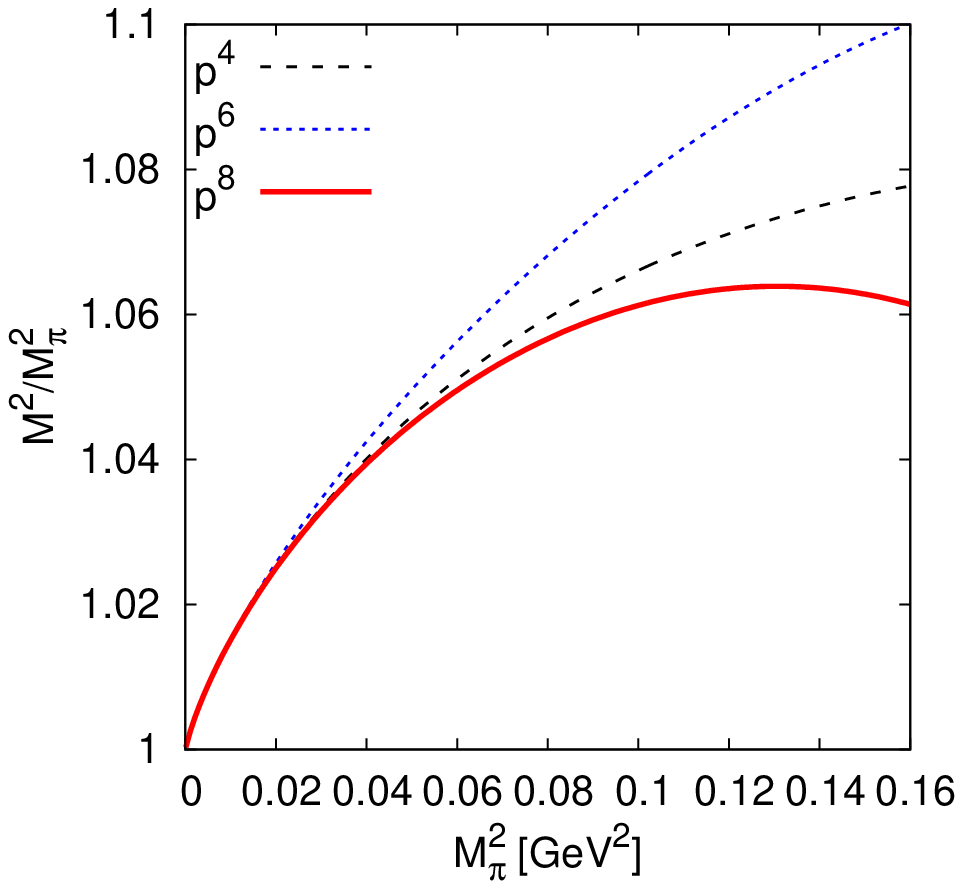}\\[-3mm]
\centerline{(b)}
\end{minipage}
\begin{minipage}{0.33\textwidth}
\includegraphics[width=0.99\textwidth]{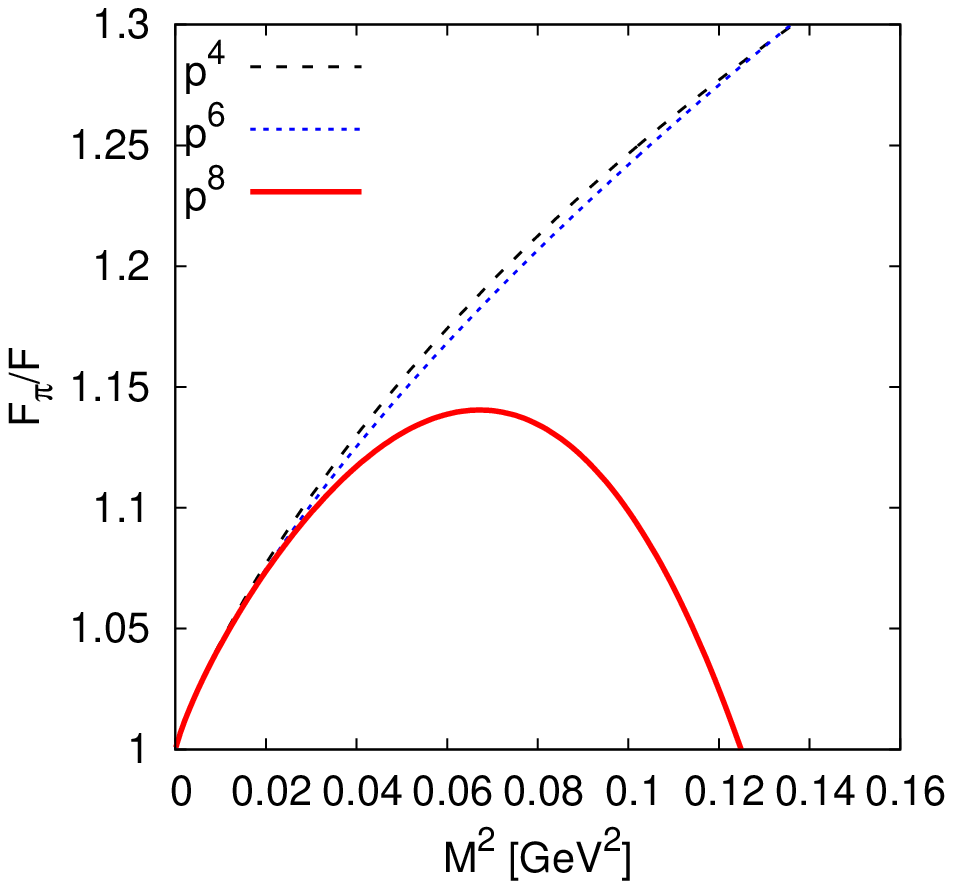}\\[-3mm]
\centerline{(c)}
\end{minipage}
\\
\sidecaption
\begin{minipage}[t]{0.33\textwidth}
\includegraphics[width=0.99\textwidth]{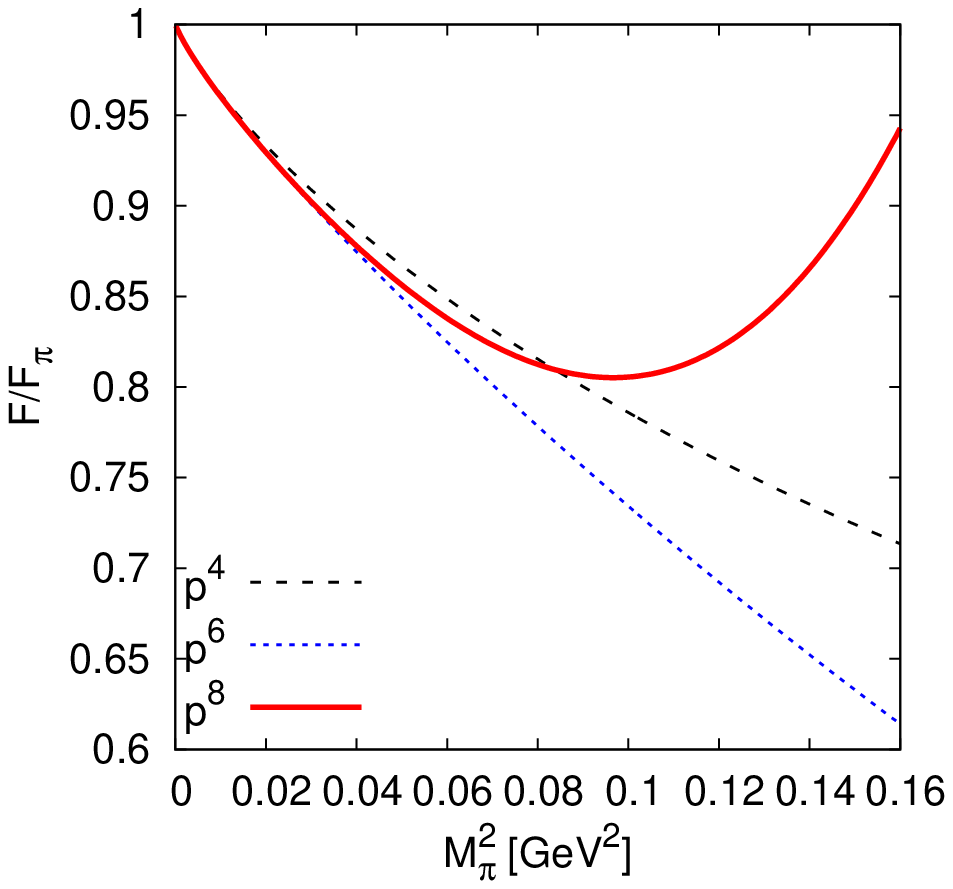}\\[-3mm]
\centerline{(d)}
\end{minipage}
\caption{(a)~The $x$-expansion for the mass, (b)~the $\xi$-expansion
for the mass, (c)~the $x$-expansion for the
decay constant, (d)~the $\xi$-expansion
for the decay constant at NLO, NNLO and NNNLO. LO is constant at 1 for all
plots. Figures from \cite{Bijnens:2017wba}.}
\label{figmassplots}
\end{figure}

The quantities (\ref{xParamEqn})--(\ref{xiParamEqn}) are shown in 
figure~\ref{figmassplots}(a--d), with the same inputs as above.
$F= 92.2/1.073$~MeV is kept constant for $x$-expansions, while
$F_\pi= 92.2$~MeV is fixed for the $\xi$-expansion. $M$ and $M_\pi$ are
varied respectively. The convergence near the physical value
$M_\pi^2\approx0.02$~GeV$^2$ is excellent.
The $\xi$-expansion converges better.

\section{ChPT for $n\bar n$-oscillations}
\label{sec:nnbar}

This work was done in collaboration with Erik Kofoed.
The baryon asymmetry of the universe is one of the unsolved problems
in particle physics. One way to solve it is via $\Delta B=2$-transitions that
appear in certain GUTs without proton decay, see~\cite{Phillips:2014fgb}
for a review and further references. There is a proposal for a free neutron
oscillation experiment at ESS in Lund that might improve the present limit
by three orders of magnitude. Neutron-antineutron oscillations need
an operator consisting of at least six quarks, schematically $uudddd$.
The lowest dimension operators of this type have dimension 9 and there are
14 of them. A classification and the short-distance running to two-loop order
is in \cite{Buchoff:2015qwa}, earlier references can be found there and
in \cite{BijnensXXX,Phillips:2014fgb}.

There are 14 dimension 9 six-quark $uudddd$ operators which transform under
chiral symmetry $SU(2)_L\times SU(2)_R$ as $(3_L,1_R)$ ($P_1,P_2,P_3$),
$(3_L, 5_R)$ ($P_5,P_6,P_7)$), $(7_L,1_R)$ ($P_4$) and the parity conjugates
$Q_1,\ldots,Q_7$. The operators $P_5,P_6,P_7$ belong to the same chiral
multiplet, so they have the same low-energy constants, while $P_1,P_2,P_3$
are not related. In fact $P_5,P_6,P_7$ are related by isospin so chiral
corrections are the same for all of these even if chiral symmetry is
spontaneously broken. $n\bar n$ transitions are isospin 1, so the operators
$P_4,Q_4$, which are isospin 3, do not contribute.

In constructing ChPT operators we use the spurion technique. The needed
spurions have two $SU(2)_L$ doublet indices symmetrized to make a $3_L$,
four $SU(2)_R$ indices symmetrized to make a $5_R$ and six $SU(2)_L$ doublet
indices to make a $7_L$ (and $L\leftrightarrow R$ for the parity-conjugates).
We use heavy baryon ChPT way to include the nucleon
field $\mathcal{N}$ with fourvelocity $v$, see e.g. \cite{Ecker:1995rk,Bernard:1995dp}. We
need to introduce also a HBCHPT antinucleon field. This is expanding
around two widely separated areas in momentum space, around $m_Nv$ for the
nucleons and $-m_Nv$ for the antinucleons, in the relativistic fields.
In HBCHPT these become independent fields. The lowest order Lagrangian
becomes\footnote{This differs slightly from what was shown during the talk
to have objects with simpler chiral transformations.}
\begin{align}
\label{LOHBCHPT}
\mathcal{L}_{LO} =\,& \frac{F^2}{4}\left\langle u_\mu u^\mu +\chi_+\right\rangle
+
\overline\NN\left(i v^\mu D_\mu + g_A u^\mu S_\mu\right)\NN
+
\overline\NNc\left(i v^\mu D_\mu - g_A u^\mu S_\mu\right)\NNc\,.
\end{align}
The definitions are the usual ChPT notation, see e.g. \cite{Ecker:1995rk}.
The fields we use are defined as
\begin{align}
\label{defN}
\NN =\,&\left(\begin{array}{c}p\\n\end{array}\right)\,,&
\NNc =\,&\left(\begin{array}{c}n^c\\-p^c\end{array}\right)\,,&
\NNt =\,&\left(\begin{array}{c}\overline{p^c}\\\overline{n^c}\end{array}\right)\,.
\end{align}
$p,n$ are the nucleon, $p^c,n^c$ the antinucleon HBCHPT fields.
The objects in $\NN^i$ in (\ref{defN}) transform under chiral symmetry all
as $\NN^i\to h(u,g_L,g_R)\NN^i$ where $h(u,g_L,g_R)$ is the usual $SU(2)_V$
compensator transformation. With these we can now construct ChPT operators
with the chiral transformation properties of the $uudddd$ quark operators.

The lowest order, $p^0$, operators are
\begin{align}
\label{LOoperators}
(3_L,1_R) :\,& R_{i_L j_L}=\left(u^\dagger\NNt\right)_{i_L} \left(u^\dagger\NN\right)_{j_L}
\nonumber\\
(3_L,5_R) :\,&R_{i_L j_L k_R l_R m_R n_R}=\left(u\NNt\right)_{k_R} \left(u\NN\right)_{l_R}
 \left(Ui\tau^2\right)_{m_R i_L}
 \left(Ui\tau^2\right)_{n_R j_L}
\end{align}
and the parity-conjugates. There is no lowest order operator for $(7_L,1_R)$.
The Dirac (fermion) indices are contracted between $\NN$ and $\NNt$.
The first operator for $(7_L,1_R)$ appears at order $p^2$:
\begin{align}
(7_L,1_R), p^2:\,&
\left(u^\dagger\NNt\right)_{i_L} \left(u^\dagger\NN\right)_{j_L}
\left(u^\dagger u_\mu ui\tau_2\right)_{k_L l_L}\left(u^\dagger u_\mu ui\tau_2\right)_{m_L n_L}
\end{align}
The operators at order $p^1$ do not contribute to $n\bar n$, at most via loops
so starting only at $p^3$.
At higher orders there are very many operators. A partial list at order $p^2$
can be found in \cite{thesisKofoed}.

\begin{figure}
\sidecaption
\includegraphics[width=0.5\textwidth]{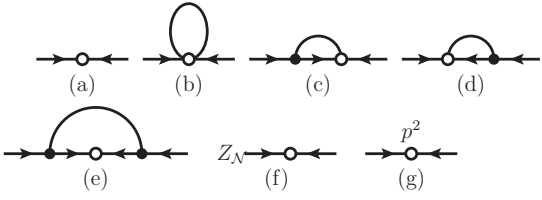}
\caption{The $n\bar n$ transition diagrams to
order $p^2$. An open dot is a vertex from the $n\bar n$ operators (\ref{LOoperators}),
a dot from the LO normal Lagrangian
(\ref{LOHBCHPT}). 
Wave-function renormalization is
indicated schematically in (f) and $p^2$ $n\bar n$-operators in (g).
A right-pointing line is a nucleon, a left-pointing line an antinucleon.}
\label{neutrondiagrams}
\end{figure}

The loop corrections can now be calculated. The diagrams are shown in
figure~\ref{neutrondiagrams}.

\begin{figure}
\sidecaption
\begin{minipage}[t]{0.33\textwidth}
\includegraphics[width=0.99\textwidth]{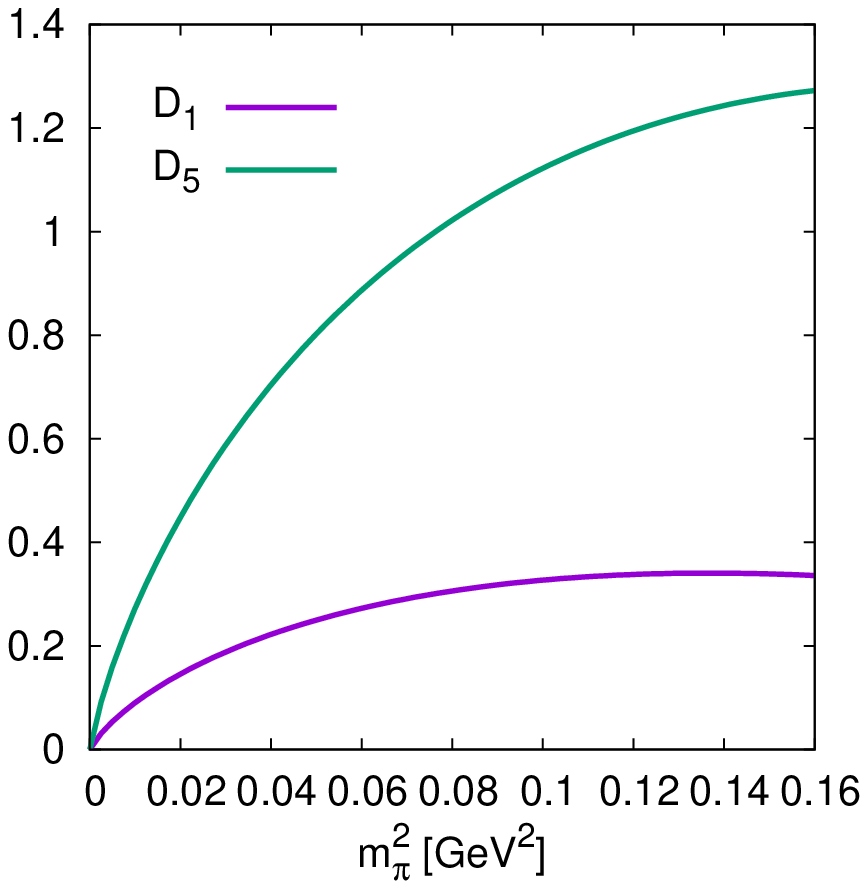}\\[-3mm]
\centerline{(a)}
\end{minipage}
\begin{minipage}[t]{0.33\textwidth}
\includegraphics[width=0.99\textwidth]{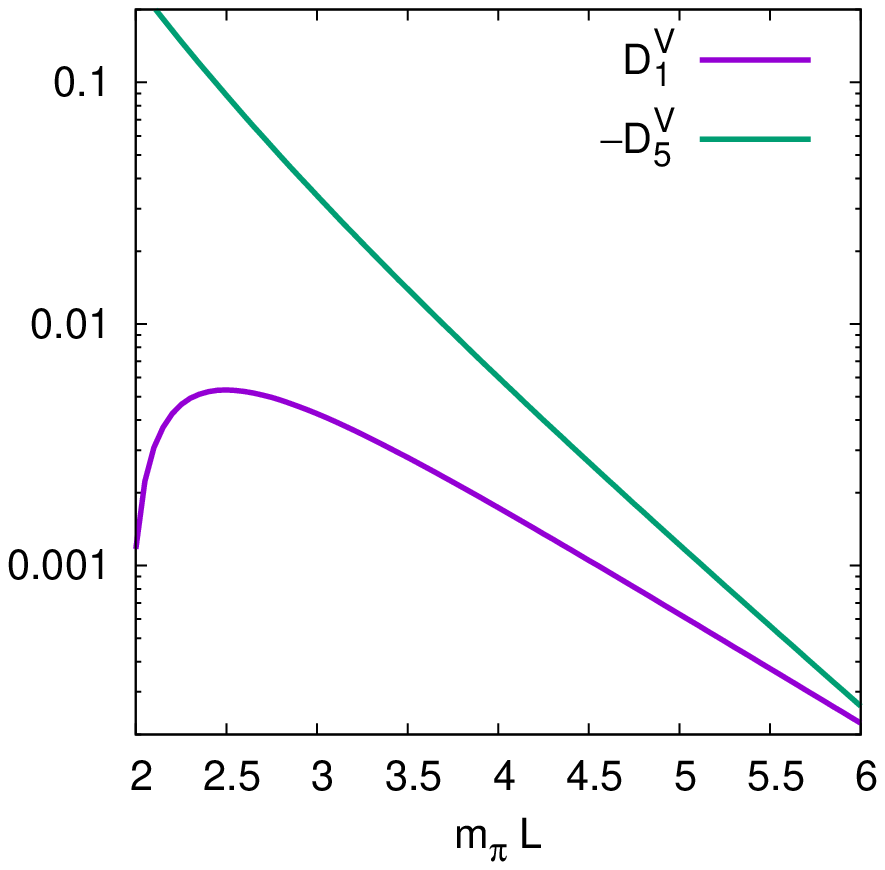}\\[-3mm]
\centerline{(b)}
\end{minipage}
\caption{\label{fignumerics} The numerical results of the pure loop contributions. (a) The infinite volume correction of (\ref{correctioninfinite})
(b) The finite volume correction. Figures from \cite{BijnensXXX}.}
\end{figure}

Due to isospin the three
$(3_L,1_R)$ and the three $(3_L,5_R)$ each have the same relative loop
corrections. The expressions for the relative corrections, i.e. multiply the
lowest order result by $1+D_i$, are
\begin{align}
\label{correctioninfinite}
D_1 =\,& \frac{m_\pi^2}{16\pi^2 F^2}\left[\left(-1-\frac{3g_A^2}{2}\right)
  \log\frac{m_\pi^2}{\mu^2}-g_A^2\right]\,,
\nonumber\\
D_5 =\,& \frac{m_\pi^2}{16\pi^2 F^2}\left[\left(-7-\frac{3g_A^2}{2}\right)
  \log\frac{m_\pi^2}{\mu^2}-g_A^2\right]\,.
\end{align}
These are plotted in figure~\ref{fignumerics}(a) for a range of $m_\pi^2$
with $F=92.2$~MeV fixed and $g_A=1.25$. Note that they are large for the
$(3_L,5_R)$ operators already at $m_\pi\approx200$~MeV.
The finite volume correction was also calculated~\cite{BijnensXXX}.
The corrections\footnote{A mistake was discovered in the numerical programs used for the plots shown during the conference, so these are different.} are shown in
figure~\ref{fignumerics}(b) for $m_\pi=135$~MeV
as a function of $m_\pi L$.

\section{Conclusions}

I discussed three different applications of ChPT that might be useful
for lattice calculations.

\section*{Acknowledgements}

This work is supported in part by the Swedish Research Council grants
contract numbers 621-2013-4287, 2015-04089 and 2016-05996 and by
the European Research Council (ERC) under the European Union's Horizon 2020
research and innovation programme (grant agreement No 668679).

\bibliography{proceedingslattice2017}
\end{document}